\documentclass[conference, twocolumn]{IEEEtran}

\usepackage{setspace}  
\usepackage[dvipdfmx]{graphicx} 
\usepackage{subfigure} 
\usepackage{tabularx}
\usepackage{arydshln}
\usepackage{umoline}
\usepackage[dvips]{color}
\usepackage{picinpar}
\usepackage{mathrsfs}
\usepackage{latexsym}
\usepackage{amssymb}
\usepackage{pifont}
\usepackage{amsfonts}
\usepackage{amsmath}
\usepackage[dvips]{color}
\usepackage{fancyhdr}
\usepackage{float}
\usepackage{cases}

\usepackage{epsfig}
\usepackage{graphics}
\usepackage[Lenny]{fncychap}
\usepackage{citesort}
\usepackage{subsections}
\usepackage{rotfloat}
\usepackage{dropping}
\usepackage{changebar}
\usepackage{url}

\begin{document}
%
%
\title{Frequency-Domain Equalization Aided Iterative Detection of Faster-than-Nyquist Signaling with Noise Whitening}
%
\author{\IEEEauthorblockN{Takumi~Ishihara and Shinya~Sugiura}\\
\IEEEauthorblockA{Department of Computer and Information Sciences\\
Tokyo University of Agriculture and Technology\\
Koganei, Tokyo 184-8588, Japan, Email: sugiura@ieee.org}  
%
\vspace*{-5mm}
\thanks{The authors are with the Department of Computer and Information Sciences, Tokyo University of Agriculture and Technology, Koganei, Tokyo 184-8588, Japan (e-mail: sugiura@ieee.org).}
\thanks{This work was supported in part by the Japan Society for the Promotion of Science (JSPS) KAKENHI Grant Number 26630170, which is gratefully acknowledged.}
}%
\markboth{}
{Shell \MakeLowercase{\textit{et al.}}: Bare Demo of IEEEtran.cls for Journals}
\maketitle
\begin{abstract}
In this paper, we propose a serially concatenated turbo-encoded faster-than-Nyquist signaling (FTNS) transceiver that takes into account FTNS-specific colored noise effects. The proposed low-complexity receiver carries out soft-decision frequency-domain equalization with the aid of the minimum-mean square error criterion while whitening the colored noise. Simulation results demonstrate that the proposed multi-stage-concatenated FTNS system achieves a better error-ratio performance than previous systems that do not consider colored noise effects {in the high-symbol-packing FTNS regime}. Furthermore, as an explicit benefit of the proposed iterative decoder, near-capacity performance is achieved with practical decoding complexity.
\end{abstract}

\begin{IEEEkeywords}
Faster-than-Nyquist signaling, frequency-domain equalization, soft-output detection, turbo coding, colored noise, minimum-mean square error 
\end{IEEEkeywords}
\IEEEpeerreviewmaketitle
\section{Introduction}
\label{sec:intro}
The concept of faster-than-Nyquist signaling (FTNS) was initially proposed by Mazo~\cite{mazo1975ftns} in the 1970s. However, FTNS has recently been rediscovered as a promising technique for next-generation wireless systems, owing to the fact that the FTNS scheme has the potential to achieve higher transmission rates than schemes based on the Nyquist criterion without imposing any bandwidth expansion~\cite{anderson2013tutrial}. 
In the conventional time-orthogonal Nyquist-criterion scenario, a symbol interval is typically set to not less than $T_\textrm{0} = 1/(2W)$, where symbols are strictly band-limited to $W$ Hz~\cite{liveris2003exploiting}. 
Under this assumption, inter-symbol interference (ISI) is not induced in the frequency-flat channel, thereby enabling ISI-free simplified detection~\cite{anderson2013tutrial}. 
In contrast, in the FTNS scheme, a symbol interval $T$ is typically set to lower than $T_\textrm{0}$, i.e., $T=\alpha T_\textrm{0}$ and $\alpha < 1$, where $\alpha$ is the packing ratio of a symbol. 
Therefore, in the FTNS scheme, more symbols are transmitted than in the conventional scheme~\cite{prlja2012reduced}. 
However, this benefit is achieved at the cost of increased ISI, which imposes a higher demodulation complexity at the receiver.

In order to eliminate the effects of ISI, several time-domain equalizers (TDEs)  have been developed for FTNS systems~\cite{liveris2003exploiting,prlja2012reduced,rusek2009multistream,mcguire2010discrete}. The TDE-based FTNS receivers are typically achieved with a high demodulating complexity for a severe ISI scenario with a low $\alpha$ and a long channel tap length. 
In order to address this problem, low-complexity hard-decision frequency-domain equalization (FDE) was applied to the uncoded FTNS scheme in \cite{sugiura2013wc}, which relies on diagonal  minimum-mean square error (MMSE) demodulation in the frequency domain.
The complexity advantage of the FDE-based FTNS receiver over its TDE counterpart is especially noticeable for the high-ISI (high-rate) scenario.
Furthermore, the hard-decision FDE of \cite{sugiura2013wc} was extended to its soft-decision (SoD) counterpart in \cite{sugiura2014tvt}, which enables practical iterative detection in the turbo-coded FTNS arrangement. 
This architecture using the powerful channel coding scheme is capable of achieving near-capacity performance, while maintaining a lower complexity, which is the chief benefit of FDE.

However, unlike their TDE counterparts, FDE-based receivers of~\cite{sugiura2013wc,sugiura2014tvt} did not take into account FTNS-specific colored noise effects~\cite{liveris2003exploiting,prlja2012reduced,rusek2009multistream,mcguire2010discrete}, and hence the FDE of~\cite{sugiura2013wc,sugiura2014tvt} may lead to performance loss. 
Most recently, motivated by \cite{sugiura2013wc}, a hard-decision FDE-based FTNS receiver that considers the effects of colored noise was proposed for uncoded FTNS systems in~\cite{fukumoto2014colored}. In this receiver, the MMSE weights are designed to  whiten the FTNS-specific colored noise, where the weight matrix is approximated to be diagonal, in order to maintain low-complexity FDE operation. 
As a result, the hard-decision FDE of~\cite{fukumoto2014colored} exhibited better BER performance than \cite{sugiura2013wc} in the uncoded FTNS scenario. 
However, since practical FTNS systems typically employ a powerful channel coding scheme~\cite{anderson2013tutrial}, such as turbo codes~\cite{sugiura2012mimo} to eliminate FTNS-specific ISI, {it is necessary to consider the iterative FTNS receiver assisted by the SoD demodulator, rather than the hard-decision version of \cite{fukumoto2014colored} for the sake of practical performance characterization of FTNS.}

Against the above-mentioned backdrop, the novel contribution of this paper is that we develop a SoD FDE-based FTNS receiver, which takes into account the effects of colored noise. 
We demonstrate that the multi-stage turbo FTNS system using the proposed noise-whitening SoD demodulator exhibits better BER performance than the previous SoD FDE scheme~\cite{sugiura2014tvt} for a low-$\alpha$ scenario, while maintaining a comparable demodulating complexity.
The remainder of this paper is organized as follows. In Section \ref{sec:system} we provide the system model of our FDE-aided FTNS system. In Section \ref{sec:sod} we propose the SoD FDE-aided FTNS demodulator that takes into consideration the colored noise, and then our three-stage-concatenated transceiver architecture is presented. Section \ref{sec:results} provides the error rate performance results of our proposed scheme. Finally, we conclude this paper in Section \ref{sec:conc}.

\section{System Model}
\label{sec:system}
In this section, we present the system model of the FDE-aided FTNS. For the sake of simplicity, we assume the additive white-Gaussian noise (AWGN) channel. 
However, this assumption will be disregarded later in Section~\ref{sec:fading}, where the frequency-selective Rayleigh fading scenario is considered. 
{Moreover, the binary phase-shift keying (BPSK) modulation is used throughout this paper, similar to \cite{sugiura2014tvt}, whereas this assumption can be readily extended to multilevel modulation schemes~\cite{matsumoto2007adaptive}.}

At the transmitter, an $M$-bit information sequence $\mathbf{b} =[b_\textrm{1},\cdots,b_\textrm{\it M}]$ $\in$ $\mathbb{Z}^\textrm{\it M}$ is modulated to $N$ complex-valued symbols $\mathbf{s} =[s_\textrm{1},\cdots,s_\textrm{\it N}]^\textrm{\it T}\in\mathbb{C}^{N}$. 
Then, a $2\nu$-length cyclic prefix (CP) is inserted at the end of the $N$ information symbols $\mathbf{s}$, and
the ($N+2\nu$)-length symbols are then passed through a shaping filter $h(t)$, where $h(t)$ may be represented by the impulse response of a root raised cosine (RRC) filter having roll-off factor $\beta$. 
Finally, the band-limited symbols are transmitted with an FTNS symbol interval $T=\alpha T_\textrm{0}$.

At the receiver, the signals matched-filtered by $h^\textrm{*}(-t)$ are given by
\begin{equation}
\setlength{\nulldelimiterspace}{0pt}
y(t)=\sum_{\it n}s_{\it n}g(t-nT) + \eta (t),
\end{equation}
where we have $g(t)=\int h(\tau)h^\textrm{*}(\tau - t)d\tau$ and $\eta (t)=\int n(\tau)h^\textrm{*}(\tau - t)d\tau$, while $n(t)$ represents a random variable that obeys the zero-mean complex-valued Gaussian distribution $\mathcal{CN}(0, N_\textrm{0})$ with noise variance $N_\textrm{0}$.
Under the assumption of perfect timing synchronization between the transmitter and the receiver, the $k$th signal sample is given by
\begin{align}
\setlength{\nulldelimiterspace}{0pt}
y_{k} &= y(kT)\\
&= \sum_{\it n}s_{\it n}g(kT-nT) + \eta (kT),
\end{align}
because $n(t)$ is convolved with $h(t)$, and the sampled noise $\eta (kT)$ is the colored noise~\cite{prlja2008receivers} with an correlation $E[\eta (mT) \eta ^\textrm{*} (nT)]=N_\textrm{0} g(mT-nT)$.

By discarding both the first and last $\nu$ samples from the $(N+2\nu )$ samples, we obtain the received signal block as follows~\cite{sugiura2013wc}:
\begin{align}
\setlength{\nulldelimiterspace}{0pt}
\hat{\mathbf{y}} &= [y_\textrm{1}, \cdots, y_\textrm{\it N}]^\textrm{\it T} \in\mathbb{C}^\textrm{\it N}\\
&= \mathbf{Gs} + \mathbf{n} \label{eq:hd-fde1},
\end{align}
where $\mathbf{G}\in\mathbb{R}^{N \times N}$ is a circular matrix, where the first column of $\mathbf{G}$ is the channel impulse responses (CIRs), whereas $\mathbf{n}$ represents the colored noise components. 
With the aid of the discrete Fourier transform (DFT)-based eigenvalue decomposition, we arrive at 
\begin{equation}
\setlength{\nulldelimiterspace}{0pt}
\mathbf{G} = \mathbf{Q^{\it T}\Lambda Q^{*}},
\end{equation}
where $\mathbf{Q}$ is the DFT matrix, the $l$th-row and $k$th-column element of which is defined by $(1/\sqrt{\it N}){\rm exp}[-2\pi j(k-1)(l-1)/N]$. Furthermore, $\mathbf{\Lambda}$ is the diagonal matrix, the diagonal elements of which are given by the fast Fourier transform (FFT) of the CIRs. Therefore, the frequency-domain counterpart of \eqref{eq:hd-fde1} may be expressed as
\begin{align}
\setlength{\nulldelimiterspace}{0pt}
\setlength\arraycolsep{10pt}
\mathbf{y}_{{\it f}}&=\mathbf{Q^{*}\hat {y}} \label{eq:hd-fde2}\\
&=\mathbf{\Lambda Q^{*}s} + \mathbf{Q^{*}n}.
\label{eq:rx-fde}
\end{align}
Here, 
the ideal MMSE weights for \eqref{eq:rx-fde} are given by~\cite{fukumoto2014colored}
\begin{equation}
\setlength{\nulldelimiterspace}{0pt}
\mathbf{W}_{\rm{colored}} = \mathbf{\Lambda}^{\it H}\left(\mathbf{\Lambda\Lambda}^{\it H}+\frac{1}{\it E_{s}}\mathbf{Q}^{*}{\it E[\boldsymbol{\eta\eta}^{H}]}\mathbf{Q}^{\it T}\right)^{-1}, \label{eq:mmse-opt}
\end{equation}
where $E[\cdot]$ represents the expectation operation.
Since the noise components $\mathbf{n}$ have an autocorrelation, as mentioned above, the matrix $\mathbf{Q}^{*} E[\boldsymbol{\eta\eta}^{\it H} ]\mathbf{Q}^{\it T}$ in \eqref{eq:mmse-opt} does not have a diagonal structure. 
This implies that the calculation of the weights $\mathbf{W}_{\rm{colored}}$ is highly complex compared with the conventional FDE~\cite{sugiura2013wc,sugiura2014tvt}. 
To reduce the high complexity imposed by the calculations of the full MMSE weights of \eqref{eq:mmse-opt}, the non-diagonal matrix $\mathbf{W}_{\rm{colored}}$ is approximated by the following diagonal matrix~\cite{fukumoto2014colored}:
\begin{equation}
\setlength{\nulldelimiterspace}{0pt}
\tilde{\mathbf{W}}_{\rm{colored}} = \mathbf{\Lambda}^{\it H}\left(\mathbf{\Lambda\Lambda}^{\it H}+\frac{N_{0}}{\it E_{s}}\mathbf{\Phi}_{\eta}\right)^{-1}, \label{eq:mmse-approx}
\end{equation}
where $E_{s}$ is the symbol power, and $\mathbf{\Phi}_{\eta}$ is a diagonal matrix of
\begin{align}
\setlength{\nulldelimiterspace}{0pt}
\mathbf{\Phi}_{\eta} &= {\rm diag}(\Phi_{\eta}[0], \ldots, \Phi_{\eta}[{\it N}-1]),
\end{align}
while we have
\begin{IEEEeqnarray}{rCL}
 \Phi_{\eta}[n] &=& \frac{1}{N}\sum_{l=0}^{N-1}\sum_{m=0}^{N-1} g((l-m)T) \nonumber\\
 && \times\exp\left({j\frac{2\pi(l-m)n}{N}}\right). \label{eq:cor1}
\end{IEEEeqnarray}
%
For comparison, the conventional MMSE weights that ignore the effects of colored noise are represented by \cite{sugiura2013wc}
\begin{equation}
\setlength{\nulldelimiterspace}{0pt}
\mathbf{W}_{\rm{white}} = \mathbf{\Lambda}^{\it H}\left(\mathbf{\Lambda\Lambda}^{\it H}+\frac{N_{0}}{\it E_{s}}\mathbf{I}_{N}\right)^{-1},
\end{equation}
where $\mathbf{I}_{N}$ is the $N$-sized identity matrix.
Finally, the time-domain symbols $\mathbf{s}$ are estimated by
\begin{equation}
\setlength{\nulldelimiterspace}{0pt}
\hat{\mathbf{s}} = \mathbf{Q}^{T}\tilde{\mathbf{W}}_{\rm{colored}}(\mathbf{\Lambda Q^{*}s + Q^{*}n}).
\end{equation}
Note that by using a sufficiently long block length $N$, the normalized overhead of $2\nu/(N+2\nu)$ imposed by the CP becomes negligible.

The uncoded FTNS scheme is not capable of approaching the near-capacity performance. Therefore, the SoD FDE-aided demodulator will be proposed for the turbo-coded FTNS system in the next section. 

\section{SoD FDE-Aided FTNS Detector and Three-Stage-Concatenated Turbo-Coded System}
\label{sec:sod}

In this section, we first propose the improved SoD FDE-based FTNS detector in the AWGN channel and extend it to the frequency-selective fading scenario. Moreover, we present the three-stage serially concatenated turbo FTNS structure using the proposed FTNS demodulator.

\begin{figure*}
\centering
\includegraphics[width=0.85\linewidth]{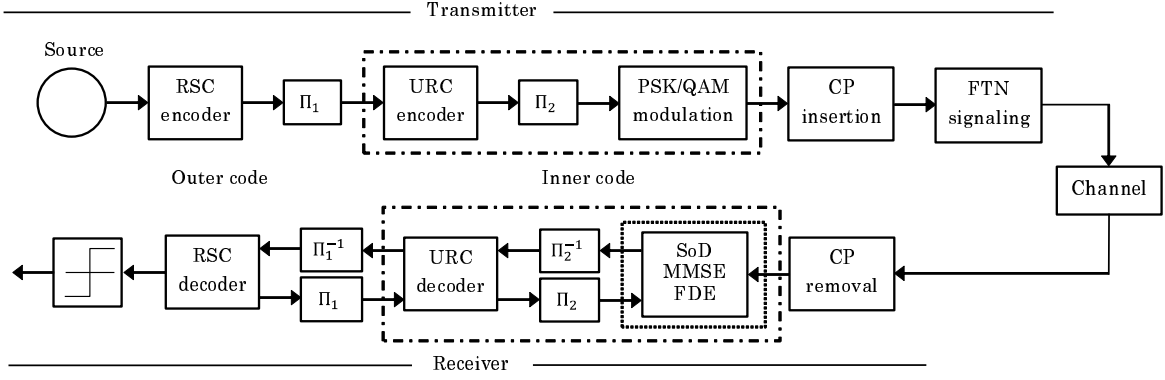}
\caption{Transmitter and receiver structures of the proposed three-stage serially concatenated FTNS architecture.  }
\label{fig:transceiver}
\end{figure*}

\subsection{Soft-Decision FDE-Aided FTNS Detection}
\label{subsec:sod}
In the SoD FDE-aided FTNS demodulator, the soft symbols $\tilde{\mathbf{s}}=[\tilde{s}_{1}, \cdots, \tilde{s}_{N}] \in \mathbb{C}^{N}$ are generated from the $\it{a\ priori}$ information that is fed back from the outer decoder. Based on the soft-interference cancellation (SIC) principle~\cite{tuchler2002apriori}, the received signals in the time-domain are given by
\begin{align}
\setlength{\nulldelimiterspace}{0pt}
\tilde{\mathbf{y}} &= \hat{\mathbf{y}} - \mathbf{G\tilde{s}} \label{eq:sod-fde1}\\
\ &= \mathbf{G(s-\tilde{s}) + n}.
\end{align}
Similar to the DFT operation of \eqref{eq:hd-fde2}, the frequency-domain counterparts of \eqref{eq:sod-fde1} are given by
\begin{align}
\setlength{\nulldelimiterspace}{0pt}
\tilde{\mathbf{y}}_{f} &= [\tilde{y}_{f, 1}, \cdots, \tilde{y}_{f, N}]^{T} \\
\ &= \mathbf{\Lambda Q^{*}(s-\tilde{s}) + Q^{*}n} \in \mathbb{C}^{N}.
\end{align}
By incorporating the colored-noise effects into the original SoD SIC-MMSE filtering~\cite{lam2007turbo}, the frequency-domain symbol estimates $\hat{\mathbf{s}}_{f}=[\hat{s}_{f, 1}, \cdots, \hat{s}_{f, N}]^{T}\in\mathbb{C}^{N}$  that are obtained by
\begin{equation}
\setlength{\nulldelimiterspace}{0pt}
\hat{s}_{f, i} = \frac{\lambda_{i}^{*}}{|\lambda_{i}|^{2}D + N_{0}\Phi_{\eta}[i]}\tilde{y}_{f, i}, \label{eq:sod-fde3}
\end{equation}
where $\lambda_{i}$ is the $i$th diagonal element of $\mathbf{\Lambda}$, whereas $D$ is the reliability value given by
$D = -\sum_{i=1}^{n} |\tilde{s}_{i}|^{2} / N$.
Note that the diagonal  approximation is imposed in \eqref{eq:sod-fde3} in order to maintain the low-complexity of FDE. Hence, this may induce a performance loss in comparison to the accurate noise whitening. Nevertheless, we demonstrate in Section~\ref{sec:results} that the proposed SoD demodulator does not lead to substantial performance loss.

Finally, the time-domain extrinsic log-likelihood ratio (LLR) outputs are calculated as follows~\cite{kansanen2007analytical}:
\begin{align}
\setlength{\nulldelimiterspace}{0pt}
\mathbf{L}_{e} &= [L_{e}(b_{1}), \cdots, L_{e}(b_{N})]^{T} \\
\ &= \frac{\gamma \tilde{\mathbf{s}} + \mathbf{Q}^{T}\hat{\mathbf{s}}_{f}}{1+\gamma\delta} \in \mathbb{R}^{N}, \label{eq:llr}
\end{align}
where we have
\begin{align}
\gamma &= \Re \left[\sum_{i=1}^{N}\frac{|\lambda_{i}|^{2} / (|\lambda_{i}|^{2}D+N_{0}\Phi_{\eta}[i])}{N}\right] \label{eq:sod-fde4}\\
\delta &= 1-D.
\end{align}
Note that the term $N_{0}\Phi_{\eta}[i]$ in \eqref{eq:sod-fde3} and \eqref{eq:sod-fde4} is replaced by $N_{0}$ in the conventional demodulator~\cite{sugiura2014tvt}.
 
\subsection{Extension to the Frequency-Selective Fading Channels}
\label{sec:fading}
Let us now extend the AWGN channel to a model applicable to frequency-selective fading channel. Consider that the delay spread associated with frequency-selective channels spans over $L_{D}T=\alpha L_{D}T_{0}$ and that the $L_{D}$ complex-valued tap coefficients are given by $q_{l} \ (l=0, \cdots, L_{D}-1)$. Then, we define the first term of (3) as follows:
$\bar{y}_{k}=\sum_{n=-\nu}^{\nu}s_{n}g(kT-nT)$.
%
The received signals may be rewritten by
\begin{align}
y_{k} &= \sum_{l=0}^{L_{D}-1}q_{l}\bar{y}_{k-l} + \eta (kT) \\
&= \sum_{l=0}^{L_{D}-1}\sum_{n=-\nu}^{\nu}s_{n}q_{l}g(kT-(l+n)T)+\eta(kT),
\end{align}
where we assume that the accurate channel impulse response is acquired at the receiver with the aid of pilot assisted channel estimation.
This system model has a circulant-matrix-based structure, in the same manner as the matrix $\mathbf{G}$ of \eqref{eq:hd-fde1} in the AWGN channel, by assuming the use of a sufficiently long CP, as compared to the effective ISI duration. Therefore, the FDE-aided FTNS operation derived in Section \ref{sec:system} is directly applicable to both the frequency-flat and frequency-selective fading scenarios.

%
%
\begin{figure*}
\centerline{
\subfigure[Proposed SoD MMSE FDE]{\includegraphics[width=0.55\linewidth]{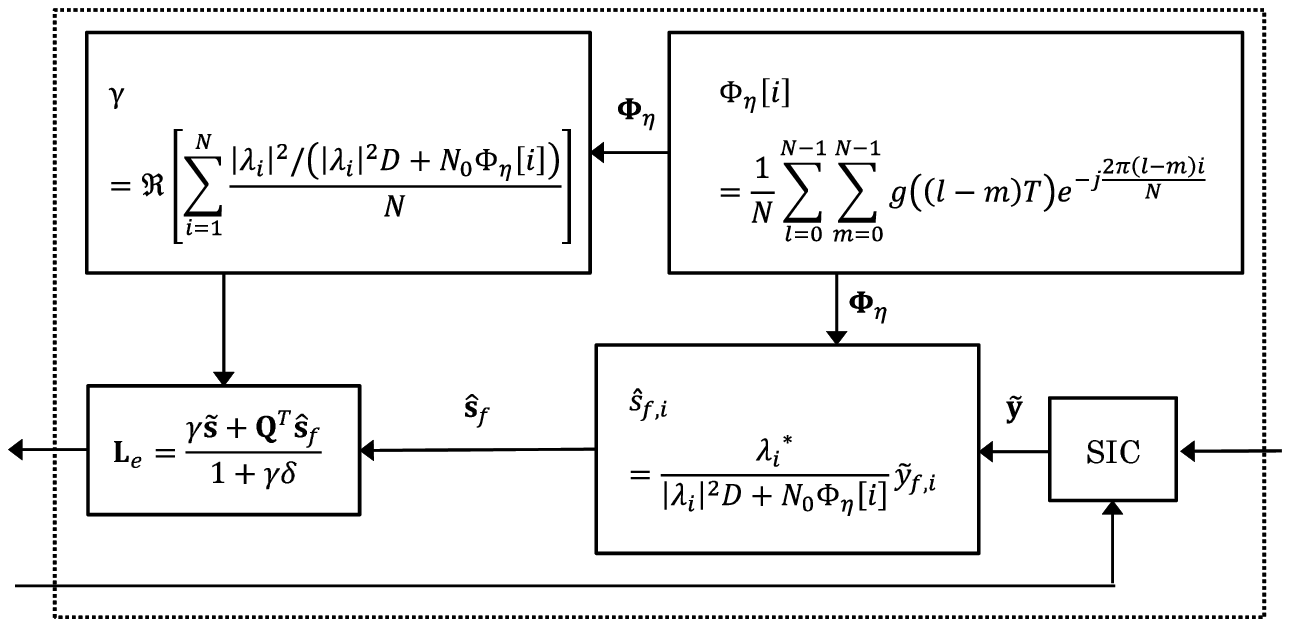}\label{fig:mmse1}}
\hfil
\subfigure[Conventional SoD MMSE FDE]{\includegraphics[width=0.38\linewidth]{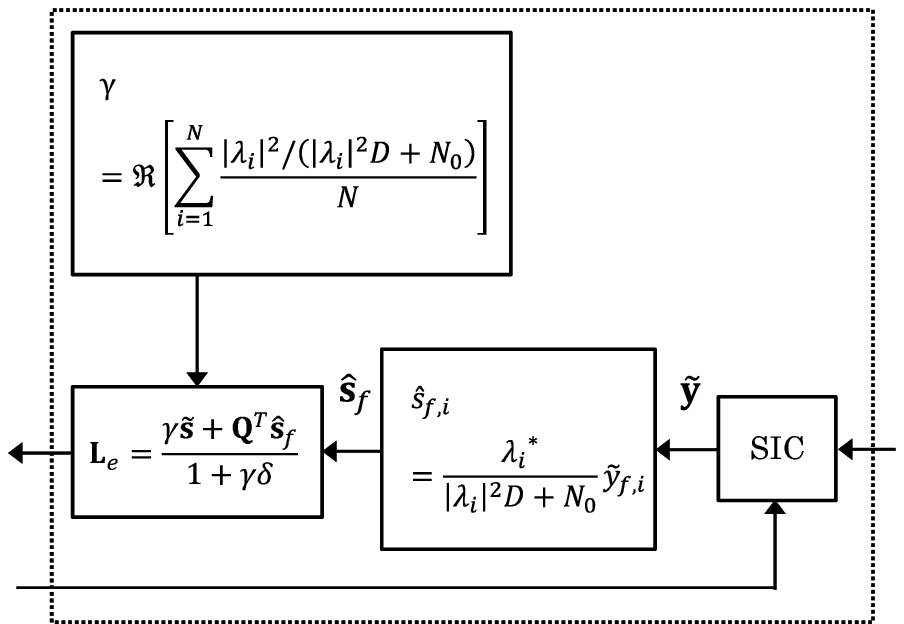}\label{fig:mmse2}}
\hfil
}
\caption{(a) Proposed SoD MMSE FDE block with noise whitening and (b) conventional SoD MMSE FDE block of \cite{sugiura2014tvt}.} 
\label{fig:mmse}
\end{figure*}

\subsection{Three-Stage-Concatenated FTNS Transceiver}
\label{subsec:3stage}
Here, we further present a multi-stage serially concatenated turbo-encoded FTNS structure~\cite{sugiura2014tvt}, which relies on the proposed low-complexity FDE-based FTNS demodulator. 
This allows us to obtain near-capacity performance, while overcoming the limitations caused by FTNS-specific ISI and colored noise. 
We use the three-stage concatenated architecture of~\cite{sugiura2014tvt}. 
At the transmitter, information bits are first encoded by the recursive systematic convolutional (RSC) encoder, which is the outer encoder. The outer encoded bits are then interleaved by the first interleaver $\mathbf{\Pi}_{1}$. Next, the interleaved bits are encoded by the unity-rate convolutional (URC) encoder~\cite{divsalar2000serial}, and the URC-encoded bits are then interleaved again by the second interleaver $\mathbf{\Pi}_{2}$. Finally, the interleaved bits are modulated to the CP-assisted FTNS of Section \ref{sec:system}.

At the receiver, three-stage iterative decoding is used, as shown in Fig.~\ref{fig:transceiver}. The three SoD decoders iteratively exchange extrinsic information in the form of LLRs. At the SoD MMSE FDE block of Fig.~\ref{fig:transceiver}, the received signals after CP removal are input, and MMSE-SIC is carried out with the aid of the extrinsic information from the URC decoder. Then, this block outputs the LLRs of \eqref{eq:llr}. 
Simultaneously, the URC decoder block receives extrinsic information from both the SoD MMSE-FDE demodulator as well as from the RSC decoder and outputs extrinsic information for both of the surrounding blocks. 
The RSC decoder of Fig.~\ref{fig:transceiver} exchanges extrinsic information with the URC decoder and outputs the estimated bits after $I_{\rm{out}}$ iterations. Here, the iterations between the SoD MMSE FDE and URC decoder blocks are referred to as inner iterations, whereas those between the URC decoder and the RSC decoder blocks are referred to as outer iterations. 
The numbers of inner and outer iterations are represented by $I_{\rm{in}}$  and $I_{\rm{out}}$ , respectively. 

The only difference between the proposed three-stage FTNS system and that of \cite{sugiura2014tvt} is the SoD MMSE FDE block of Fig.~1. In order to elaborate further, Fig.~\ref{fig:mmse1} shows the proposed SoD MMSE FDE diagram of Section \ref{subsec:sod}, whereas Fig.~\ref{fig:mmse2} shows the conventional counterpart. In Fig.~~\ref{fig:mmse1}, the estimated symbols $\hat{s}_{f}$ and $\gamma$ are calculated considering the impact of colored noise, whereas Fig.~~\ref{fig:mmse2} assumes white noise.

\subsection{Evaluation of the Demodulation Complexity}
\label{subsec:complexity}
In this section, we roughly evaluate the computational complexity of the proposed SoD FDE-aided FTNS demodulator. First, as mentioned above, the difference in the improved SoD FDE-aided scheme, as compared to the previous scheme~\cite{sugiura2014tvt}, is the presence of $\Phi_{\eta}[n]$ in \eqref{eq:cor1}, which is added for the purpose of whitening the colored noise. 
Here, the calculation of $\Phi_{\eta}[n]$ requires $2N^{2}$ real-valued multiplications. Therefore, the total calculation cost in (11) is $2N^{2}$ real-valued and $N$ complex-valued multiplications.

Importantly, $\Phi_{\eta}[n]$ remains constant for each of the FTNS parameters $(\alpha, \beta)$. 
More specifically, $\Phi_{\eta}[n]$ depends only on the shaping filter $h(t)$ and the block length $N$. 
Hence, the calculations of $\Phi_{\eta}[n]$ in \eqref{eq:cor1} are stored in the memory of the receiver in advance of the demodulation. 
This means that in the proposed demodulator, only an additional $N$ complex-valued multiplications is required for the calculation of $N_{0}\mathbf{\Phi}_{\eta}$ in (19) and \eqref{eq:sod-fde4}.

Furthermore, $\Phi_{\eta}[n]$ of \eqref{eq:cor1} is unaffected the presence of either ISI or fading. Hence, the demodulation complexity of the proposed receiver in the frequency-selective fading channel remains almost unchanged from the conventional demodulator.

\section{Simulation Results}
\label{sec:results}
To characterize the achievable performance of the proposed FDE-aided FTNS system, we calculated the BER using Monte Carlo simulations.
We considered two benchmark schemes, i.e., the conventional FDE-aided FTNS systems of \cite{sugiura2013wc,sugiura2014tvt} and the Nyquist-criterion-based system, employing a maximum a priori demodulator. 
Here, we assumed the use of an RRC filter having a roll-off factor of $\beta = 0.5$. Moreover, we considered the transmissions of BPSK-modulated symbols. 
Finally, the number of outer and inner iterations was set to $I_{\rm{out}}=21$ and $I_{\rm{in}}=2$.

\begin{figure}
\centering
\includegraphics[width=0.95\linewidth]{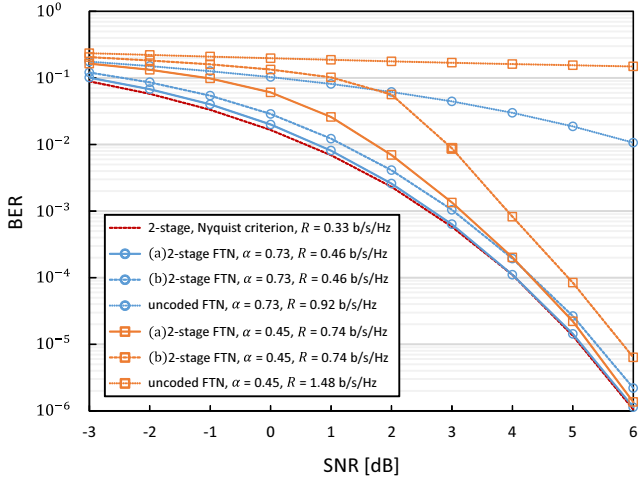}
\caption{Achievable BERs of the proposed FDE-aided two-stage RSC-encoded FTNS systems using (a) the proposed SoD FDE-based demodulator and (b) the conventional counterpart \cite{sugiura2014tvt}. The FTNS parameters were set to $(\alpha, \beta,\nu)=(0.45,0.5,10)$ and $(0.73,0.5,10)$.} 
\label{fig:2stage_BER}
\end{figure}
In Fig.~\ref{fig:2stage_BER} we calculated the achievable BER of the two-stage RSC-coded  iterative FTNS systems, having the parameter sets of $(\alpha,\beta,\nu)=(0.45, 0.5, 10)$ and $(0.73, 0.5, 10)$. 
Furthermore, we also plotted the BER curves of the uncoded FTNS system using the hard-decision FDE of \cite{fukumoto2014colored}. 
We used the half-rate RSC$(2,1,2)$ code having the octal generator polynomials of $(3,2)$, and the block length was set to $N=2^{17}=131,072$ for the turbo-coded systems. 
Note in Fig.~\ref{fig:2stage_BER} that the BER of the proposed FDE-based iterative FTNS system converged to the Nyquist-criterion-based limit. This ensures that the proposed demodulator does not impose any substantial performance loss, despite the diagonal approximation used in the MMSE-weight derivation \eqref{eq:mmse-approx}.
Furthermore, whereas in a lower FTNS rate scenario of 0.46 bps/Hz, the proposed FDE-based FTNS system and that of \cite{sugiura2014tvt} exhibited similar performances, the performance advantage of the proposed FDE-based FTNS system was clear for a higher-rate FTNS scenario of 0.74 bps/Hz.

\begin{figure}
\centering
\includegraphics[width=0.95\linewidth]{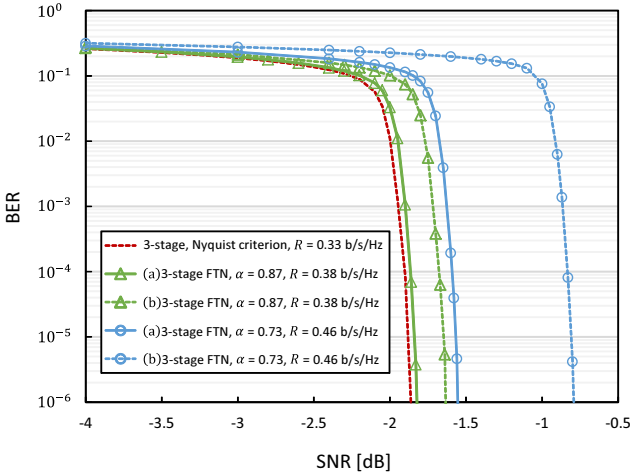}
\caption{Achievable BERs of the proposed FDE-aided three-stage FTNS systems using (a) the proposed SoD FDE-based demodulator and (b) the conventional counterpart \cite{sugiura2014tvt}.  The FTNS parameters were set to $(\alpha,\beta,\nu)=(0.73,0.5,10)$ and $(0.87,0.5,10)$.} 
\label{fig:3stage_BER}
\end{figure}
Next, in Fig.~\ref{fig:3stage_BER}, we plotted the achievable BER performances of the RSC- and URC-encoded three-stage FTNS systems, using the FTNS parameter sets of $(\alpha,\beta,\nu)=(0.73, 0.5, 10)$ and $(0.87,0.5,10)$. 
Note that in Fig.~\ref{fig:3stage_BER}, as an explicit benefit of the three-stage turbo architecture~\cite{sugiura2012mimo}, the BER curves of all of the systems considered exhibited turbo cliffs.
Similar to Fig.~\ref{fig:2stage_BER}, whereas the proposed FDE-based FTNS receiver exhibited a comparable BER performance to the conventional FDE-based FTNS receiver \cite{sugiura2014tvt} for a lower-rate FTNS scenario of 0.38 bps/Hz, the performance difference became clear upon increasing the transmission rate to 0.46 bps/Hz.
However, the increase in performance of the proposed demodulator over the conventional demodulator \cite{sugiura2014tvt} was smaller in the three-stage FTNS architecture, in comparison to that shown in the two-stage counterpart of Fig.~\ref{fig:2stage_BER}.

\begin{figure}
\centering
\includegraphics[width=0.95\linewidth]{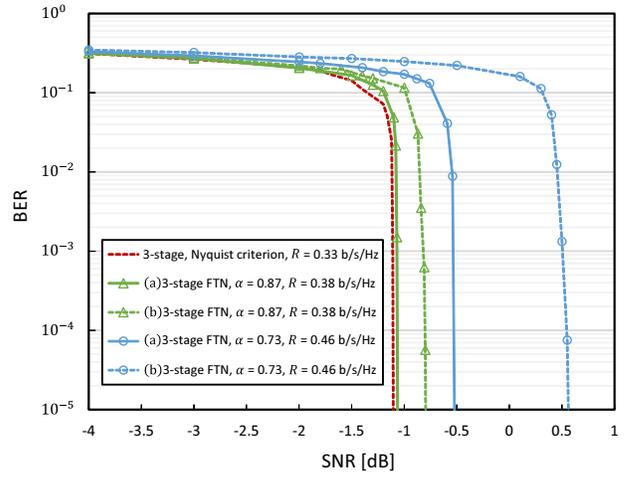}
\caption{Achievable BERs of the proposed FDE-aided three-stage FTNS systems using (a) the proposed SoD FDE-based demodulator and (b) the conventional counterpart \cite{sugiura2014tvt}.  The FTNS parameters were set to $(\alpha,\beta,\nu)=(0.73,0.5,24)$ and $(0.87,0.5,24)$.} 
\label{fig:3freq_BER}
\end{figure}
Fig.~\ref{fig:3freq_BER} shows the BER performance of the proposed three-stage-concatenated FTNS systems under frequency-selective block Rayleigh fading environments. A constant block-length of 512 bits and a constant CP-length of $2\nu=48$ were considered. The delay spread was set to $L_{D}=20$, and the fading tap coefficients $q_{l}$ were randomly generated according to the complex-valued Gaussian distribution $\mathcal{CN}(0,1/L_{D})$. Moreover, we using an interleaver length of $2^{17}$ and the FTNS parameter sets of $(\alpha,\beta,\nu)=(0.73,0.5,24)$  and $(0.87,0.5,24)$. 
The performance advantage of the proposed scheme is shown in Fig.~5, in a similar manner to performance advantage of the AWGN scenario shown in Fig.~4.

\begin{figure}
\centering
\includegraphics[width=\linewidth]{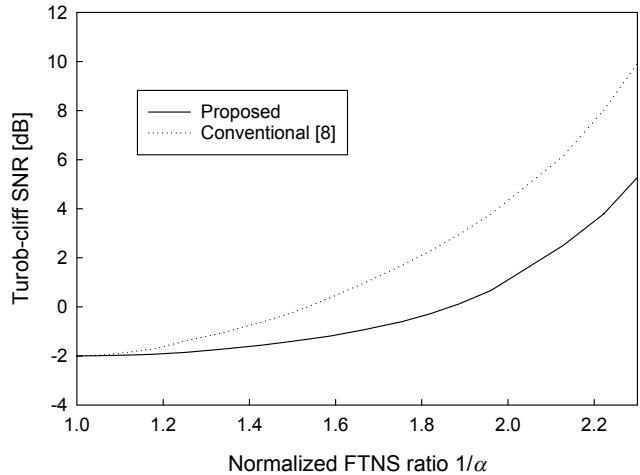}
\caption{The effects of the normalized FTNS ratio $1/\alpha$ on the error-free SNR of the three-stage FTNS schemes considered in Fig.~\ref{fig:3stage_BER}, where the parameters were set to $(\beta,\nu)=(0.5,10)$, while assuming the AWGN channel.} 
\label{fig:cliff-snrs}
\end{figure}
{Finally, in Fig.~\ref{fig:cliff-snrs} we further evaluated the performance gap between the proposed and the conventional FDE-aided FTNS receivers, by investigating the effects of the normalized FTNS ratio $1/\alpha$ on the error-free SNR of the three-stage FTNS schemes, where the parameters were set to $(\beta,\nu)=(0.5,10)$. 
Here, the AWGN channels were assumed, similar to Fig.~\ref{fig:3stage_BER}, and the error-free SNR was defined by that corresponding to BER$= 0.01$, which tends to be a slightly lower SNR than that of the turbo cliff.
It was clearly shown that upon increasing the normalized FTNS ratio $1/\alpha$ from 1 to 2.3, the performance gap of the proposed receiver over the conventional one~\cite{sugiura2014tvt} increased. This is because the effects of colored noise becomes more explicit for a lower-$\alpha$ (higher-rate) FTNS scenario, hence exhibiting the benefit of the proposed receiver.}

\section{Conclusions}
\label{sec:conc}
In the present paper, we proposed an SoD FDE-aided serially concatenated FTNS architecture that takes into account FTNS-specific colored noise effects. With the aid of the approximated MMSE weights, the proposed detector is capable of effectively compensating for colored noise while eliminating ISI imposed by FTNS. Simulation results demonstrated that the proposed FTNS system outperforms the previous scheme in a high-rate FTNS scenario.

\section*{Acknowledgment}
This work was supported in part by the Japan Society for the Promotion of Science (JSPS) KAKENHI Grant Number 26630170, which is gratefully acknowledged.


\IEEEtriggeratref{2}
\bibliographystyle{IEEEtran}
\bibliography{mybibfile}

\end{document}